# Interlaboratory comparison of particle filtration efficiency testing equipment

**Timothy A. Sipkens[1], Ruth Perez Calderon[1], Richard G. Green[1], Andrew Oldershaw[1], Gregory J. Smallwood[1]**

[1] *Metrology Research Centre, National Research Council Canada, Ottawa ON Canada*

**Abstract**

This work presents the results of two interlaboratory comparisons of particle filtration efficiency measurements performed by a network of laboratories across Canada and Australia. Testing across multiple layers of a common verification material demonstrates a constant size-resolved quality factor when layering uncharged materials. Size-resolved filtration curves also match expectations, with increasingly size-dependent curves and a predictable increase in the PFE. Candidate reference materials with controlled material properties were also tested across multiple laboratories. Each set of materials sharing a common charge level show specific trends with the material basis weight. Respirators showed more consistency between the laboratories than the other filters. However, across a majority of the tests, dark uncertainties, which are otherwise unexplained variability between laboratories, are significant. This leaves room to improve the test method by developing improved verification procedures and additional reference materials.

## 1. Introduction

Face masks, respirators, and other related products are essential personal protective equipment, which protect against a range of respiratory hazards, including infectious pathogens, forest fire smoke, and occupational hazards. Amongst the fundamental quantities characterizing these products is particle filtration efficiency (PFE) that quantifies the ability of a filter to capture particles in the air. The considerable attention given to PFE as a result of the COVID-19 pandemic highlighted the need for a better understanding of test results. Quantifying the uncertainty of test results, as well as the precision and bias of the associated test methods, are key to removing doubt as to the transferability or traceability of the test methods used to determine PFE. Given potential challenges with understanding dark uncertainties reference materials supported by robust interlaboratory comparisons, present a more viable pathway to establishing traceability.

Interlaboratory comparisons (ILCs) to quantify the reproducibility of measurements of respiratory protection are not widely available and are often very limited in the range of materials and participating laboratories. Li et al. [1] considered two electret materials and one fiberglass sample, making size-resolved measurements across two laboratories. Bourrous et al. [2] performed an ILC with three laboratories using filtration media making up community face coverings, with the goal of identifying optimal materials for use in such products. In that case, the aerosols varied between the three laboratories (NaCl, DEHS, and PSL), which were size-selected using an aerodynamic particle sizer. While impaction is the dominant mechanical filtration mechanism and is related to aerodynamic diameter, electrostatic filtration may differ between the different particle types and could expand uncertainties beyond simple interlaboratory differences. Tessarolo et al. [3] presented a comparison of measurement of bacterial filtration efficiency (BFE) and pressure drop by six non-accredited laboratories in Italy. This study spanned three separate types of masks and provided some decomposition of the variability based on structure in the measurements. Fouqueau et al. [4] presented a comparison across four laboratories examining the difference between particle filtration efficiency at 3 μm (two laboratories) and bacterial filtration efficiency (the remaining two laboratories).

This work aims to rigorously quantify differences between measurements of PFE across a series of independent laboratories, specifically for the sodium chloride test typically used for testing respirators. It expands greatly on previous comparisons, with a substantially wider range of filters and a larger number of laboratories spanning two countries, allowing for a more robust estimate and understanding of interlaboratory variability in respiratory protection testing. Results are supplemented with measurements of pressure resistance, enabling discussion of filter quality [5], and are examined in terms of trends in certain underlying material properties, including basis weight (using qualified categories), charge state, number of layers, and environmental conditioning (specific to a single respirator).

# 2. Methods

## 2.1 Experimental protocol

All of the measurements throughout this work were conducted using a challenge aerosol consistent with many standards for particle filtration testing of respirators such as that used in NIOSH TEB-APR-STP-0059 [6] for certifying N95 respirators, GB 2626-2019 [7] for certifying KN95 respirators, and CSA Z94.4.1:21 [8] for certifying CAN95 respirators. Sodium chloride is nebulized to form a challenge aerosol having a count median mobility diameter (CMD) of 75 nm ± 20 nm and a geometric standard deviation (GSD) less than or equal to 1.86. In accordance with the cited test methods, the particles are typically dried, diluted, and neutralized.

A range of instruments were permitted to measure aerosol concentrations and are separated into three categories. Due to the prevalence of the TSI 8130A, both in terms of its explicit mention in the TEB-APR-STP-0059 test method [6] and in terms of its use by participating laboratories, the instrument receives its own category. The remaining instruments are categorized either as photometer-based, where scattering is used to measure the quantity of particles, or count-based, where particles are counted (e.g., the Particle Filtration Efficiency Measurement System (PFEMS), which was developed by the National Research Council Canada [9, 10]). Note that the TSI 8130A (TSI, USA) is a photometer-based instrument, even though it has its own category. Other photometer-based instruments include the DustTrak (TSI 8540/8543, USA), PMFT 1000 (Palas, Germany), and 100X (ATI, USA). Measurements with a photometer act as a close surrogate to the mass of particulate captured by a filter [11, 12]. As such, those laboratories employing count-based measures were instructed to convert their measurements to a mass-equivalent result for comparison against other data using an appropriate method [11], with differences in these representations having been measured in the literature previously [1, 11, 13].

Within this framework, measurements were conducted across two interlaboratory comparisons, summarized in Table 1. The first ILC involved laboratories from Canada that took place in 2021, while the second ILC was a bilateral comparison involving laboratories from both Canada and Australia that took place in 2022. The distinction between the first and second ILC is otherwise unimportant and receives little attention in the remainder of this work. Rather, throughout, the measurements are separated into *components*, used to group a series of subcomponents investigating a particular set of samples (e.g., measurement of a fiberglass material that is layered). In all cases, internal measurements made by the National Research Council Canada prior to sample distribution indicated uniformity amongst the samples sufficient to consider the distributed samples as randomly selected from the sample population.

Laboratory identifiers throughout this manuscript are anonymized and are identified on a per-ILC basis (i.e., lab 1 for the first ILC is not the same as lab 1 for the second ILC). For the first ILC, laboratories are numbered according to results from measurements for one layer of Green Line Paper (prescribed calibration media documented in the TEB-APR-STP-0059 test method [6]). For the second ILC, laboratories are sorted by the median of the results from the three sets of measurements in the fourth component, described in Sec. 2.1.2.

**Table 1.** Summary of the various components of the two interlaboratory comparisons presented in this work. The number of subcomponents refers to the number of different cases that were considered within each component (e.g., measuring 1 to 5 layers for the first component of the first ILC).

| ILC | Component | No. of subcomponents | Sample type | Target parameters | No. of laboratories |
|---|---|---|---|---|---|
| 1st | 1 | 5 | Green Line Paper (TSI 813010) | $\Delta p$, PFE, layering | 16 |
| 1st | 2 | 10 | Candidate reference materials | $\Delta p$, PFE, matl. charge state | 16 |
| 1st | 3 | 2 | Respirator | $\Delta p$, PFE, initial/minimum PFE | 12 |
| 2nd | 4 | 3 | Respirator | $\Delta p$, PFE | 15 |

$\Delta p$: pressure resistance; $P$: penetration

### *2.1.1 First ILC*

The first ILC had components driven by differences in the samples being tested, which spanned different materials, construction, and charge states. There are three components relevant here:

**Component 1.** This first component examined the PFE of layered Green Line Paper (TSI 813010), as supplied from a single lot by the National Research Council Canada. Green Line Paper is typically supplied for verification of TSI 8130A instruments [6]. The material is composed of uncharged fiberglass and exhibits a strongly-size dependent PFE relative to many other materials. The filtration media was tested flat, i.e., without bending or cupping, and at a face velocity of 10 cm $s^{-1}$. Tests differed in terms of layering multiple sheets of the filtration media, from 1 to 5 in increments of 1. This allows for some assessment of layering and allows for spanning multiple PFE levels, while holding the other properties of the material constant.

**Component 2.** PFE was measured on a set of candidate reference materials composed of flat melt-blown polypropylene sheets manufactured by Roswell Downhole Technologies and having some control of the underlying material properties. These materials were also used in Sipkens et al. [11] and span a known range of PFE levels, basis weights, and charge states for a similar class of material. While the specifics of the material synthesis are outside of the scope of this work, the outcome is particle filtration media with PFE values that span 40% – 99%, with a qualified level of charging (none, half, and full). The first six subcomponents are ordered according to the expected PFE, as measured by the National Research Council Canada prior to sample distribution. This also corresponds to ordering the samples with respect to increasing charge state, with the first two subcomponents having samples with no charge, the third subcomponent having a sample with half of the applied charge, and the remainder of the subcomponents having materials with a full charge applied. The final four subcomponents correspond to pairs of materials each from the same lot (thus having the same basis weight and charge state) that were tested both as-received (unconditioned) and environmentally conditioned according to the NIOSH TEB-APR-STP-0059 test method, namely for 25 hr ± 1 hr at a temperature of 38 °C ± 2.5 °C and at a relative humidity of 85 % ± 5 % RH. Samples were tested at a face velocity of 10 cm $s^{-1}$.

**Component 3.** PFE was measured on formed respirators that were supplied to each of laboratories from a single lot. Measurements were recorded both in terms of minimal loading (i.e., initial PFE) and the minimum PFE as the products are loaded, consistent with requirements in respirator testing. Samples were tested at a flow rate of 85 LPM.

### *2.1.2 Second ILC (bilateral)*

The second ILC had a focus on the test method prescribed in CSA Z94.4.1:21 [8]:

**Component 4.** For this component, PFE was measured on formed respirators, with measurements reported at the minimum loading possible, and were generally measured following the instructions in CSA Z94.4.1:21 [8]. There was some variability in terms of the minimum amount of loading at which a measurement could be reported, given the range of instruments used in this study. Conditioned samples were tested in both inhalation and exhalation configurations, while unconditioned samples were tested in the inhalation configuration, as described in CSA Z94.4.1:21. Conditioning of samples is performed following the guidelines set out in CSA Z94.4.1:21 (which match the requirements in NIOSH TEB-APR-STP-0059). Samples were tested as soon as possible upon removal from the conditioning environment or placed and sealed in a gas-tight container at ambient conditions, and tested within 10 hours.

## 2.2 Log-penetration as a surrogate for PFE

Results for PFE are all reported in terms of the logarithm of penetration, with the equivalencies in terms of penetration and percent PFE shown in Table 2, for reference. This treatment is expected to be more representative of the distribution of results as a consequence of how penetration scales with material thickness and other underlying properties [5] and the scaling of the noise with penetration [14], though this will be revisited in the results. This is also the approach used in a number of academic works (e.g., [15]) and is cited in ASTM F2299 [16]. While uncertainties may not perfectly scale with the logarithm of penetration (as will be shown), this is a far better approximation than linear scales with respect to PFE, where standard deviations can easily extend above 100% for PFE, largely making this an unacceptable metric. Such a treatment is also encoded in the filter quality factor [5], a metric balancing the penetration and pressure resistance that yields a quantitative measure of the performance of a sample:

$$Q_\mathrm{F} = \frac{-\log_{10} P}{\Delta p} = \frac{-\log_{10}\left(1-\eta/100\%\right)}{\Delta p} \tag{1}$$

where $Q_\mathrm{F}$ is the quality factor, $P$ is the penetration, $\Delta p$ is the pressure resistance, and $\eta$ is the PFE as a percent. This quantity has been shown to be particularly useful in distinguishing between different materials, e.g., [15, 17-21]. Note that the quality factor is often stated in terms of the natural logarithm, such that this must be verified when comparing to other works. The $\log_{10}$ form is chosen here due to convenience of the corresponding PFEs for integer values, e.g., $\log_{10}P = -1$ corresponds to 90% PFE. It is also worth noting that the quality factor requires a consistent face velocity and particle size, which further restricts comparability between studies but makes the criterion useful if the test is identical across different studies. It is expected that a material, when layered will have a constant quality factor as the ratio of log-penetration and $\Delta p$ does not change. This relationship is linear on a log-penetration–$\Delta p$ plot, with quality factor controlling the slope of the relationship.

Plots in this work typically reverse the penetration axis such that PFE still increases as one moves up the corresponding panels. Uncertainties in log-penetration will result in skewed uncertainties in terms of the original penetration and PFE. Standard deviations stated in terms of log-penetration will correspond to a geometric standard deviation in terms of the penetration itself. From simple error propagation, absolute uncertainties in PFE and penetration will be the same with opposite skew [14].

**Table 2.** Reference table showing the logarithm (base-10) of the penetration, the penetration as a fraction, and PFE as a percentage.

| $\log_{10}P$ | −3.000 | −2.699 | −2.301 | −2.000 | −1.699 | −1.523 | −1.301 | −1.000 | −0.301 | 0 |
|---|---|---|---|---|---|---|---|---|---|---|
| Penetration, $P$/- | 0.001 | 0.002 | 0.005 | 0.01 | 0.02 | 0.03 | 0.05 | 0.1 | 0.5 | 1 |
| PFE, $\eta$/% | 99.9 | 99.8 | 99.5 | 99 | 98 | 97 | 95 | 90 | 50 | 0 |

## 2.3 Statistical method: Hierarchical random effects model

Data is analyzed according to a random effects model. In this framework, effects are perceived as random variables, holding some distribution. Then, the measured value, $y_{jk}$, is modeled as a combination of effects

$$y_{ij} = \mu_i + l_{ij} + \varepsilon_{ij} \tag{2}$$

where $\mu_i$ is the consensus value for the $i$th test; $l_{ij}$ is the random effect for the $j$th laboratory and represents dark uncertainties; and $\varepsilon_{ij}$ is the remaining error within a given laboratory. The laboratory effect, $l_{ij}$, is assumed to be normally-distributed and unbiased, $l_{ij} \sim N(0, \sigma_{l,i}^2)$, and is conditioned on the tests, allowing for different degrees of lab-to-lab variation for each test. Here, the variance $\sigma_{l,i}^2$ is used to represent dark, interlaboratory uncertainties or biases [22], which may not be evident in measurement by a single laboratory (i.e., are *dark* or hidden) and are representative of reproducibility. Errors, $\varepsilon_{ij}$, are taken as normally distributed with a variance from the repeat measurements made by each laboratory (i.e., Type A uncertainties or repeatability).

The distribution for the effects and consensus value were determined using Markov Chain Monte Carlo (MCMC). This approach generates a distribution for the interlaboratory variability, $l_{ij}$, and the consensus value. To minimize the MCMC burn-in period, the set of $l_{ij}$ were initiated about the average for each quantity and $\mu_i$ about the global mean. To restrict the solution space and improve convergence with this Bayesian approach, we also apply priors (encoding approximate information known before the statistical analysis) to these quantities. The prior on $\mu_i$ was taken as exponential about the global average, maximizing the information entropy when only using the mean. The prior on the standard deviation of $l_{ij}$ was taken as half Cauchy distributions with a median of the difference between the global mean and the laboratory mean. The method used here makes use of the Just Another Gibbs Sampler (JAGS) code (Hornik et al., 2003), analogous to the approach used in Ref. [23].

In running the statistical analysis, outliers are typically not removed, so as to capture the broader variability possible in these measurements. This includes in determining the consensus value and the magnitude of the effects. However, when discussing physical trends, outliers are identified using the generalized extreme Studentized deviate (GESD) test, typically using a threshold factor of 0.6. Exclusion of such measurements would act to underestimate the dark uncertainties.

# 3. Results

## 3.1 Green Line Paper and layering

Figure 1 demonstrates PFE measurements made by each laboratory, for 1 to 5 layers of Green Line Paper. Results suggest that the log-penetration transformation is likely reasonable, with roughly normally-distributed data following the transformation for most cases. Results still contain some heteroskedasticity in terms of log-penetration, with a wider spread in the results for higher PFE. It is worth noting, however, that plotting the results on a linear scale yields far more heteroskedasticity, particularly when considering interlaboratory variability, further affirming the choice to use log-penetration. The remaining heteroskedasticity could be caused by a number of effects, including: (1) the fact that these are integrated penetrations rather than size-resolved penetrations, such that these results are an additional step removed from the original filtration mechanisms and thus not perfectly aligned with the model discussed by Hinds [5]; (2) a combination of Poisson and other detector noise in the photometers [14] that don't scale with the filtration mechanisms; or (3) secondary interfacial effects between the layers (though size-resolved measurements discussed subsequently suggest that this is unlikely).

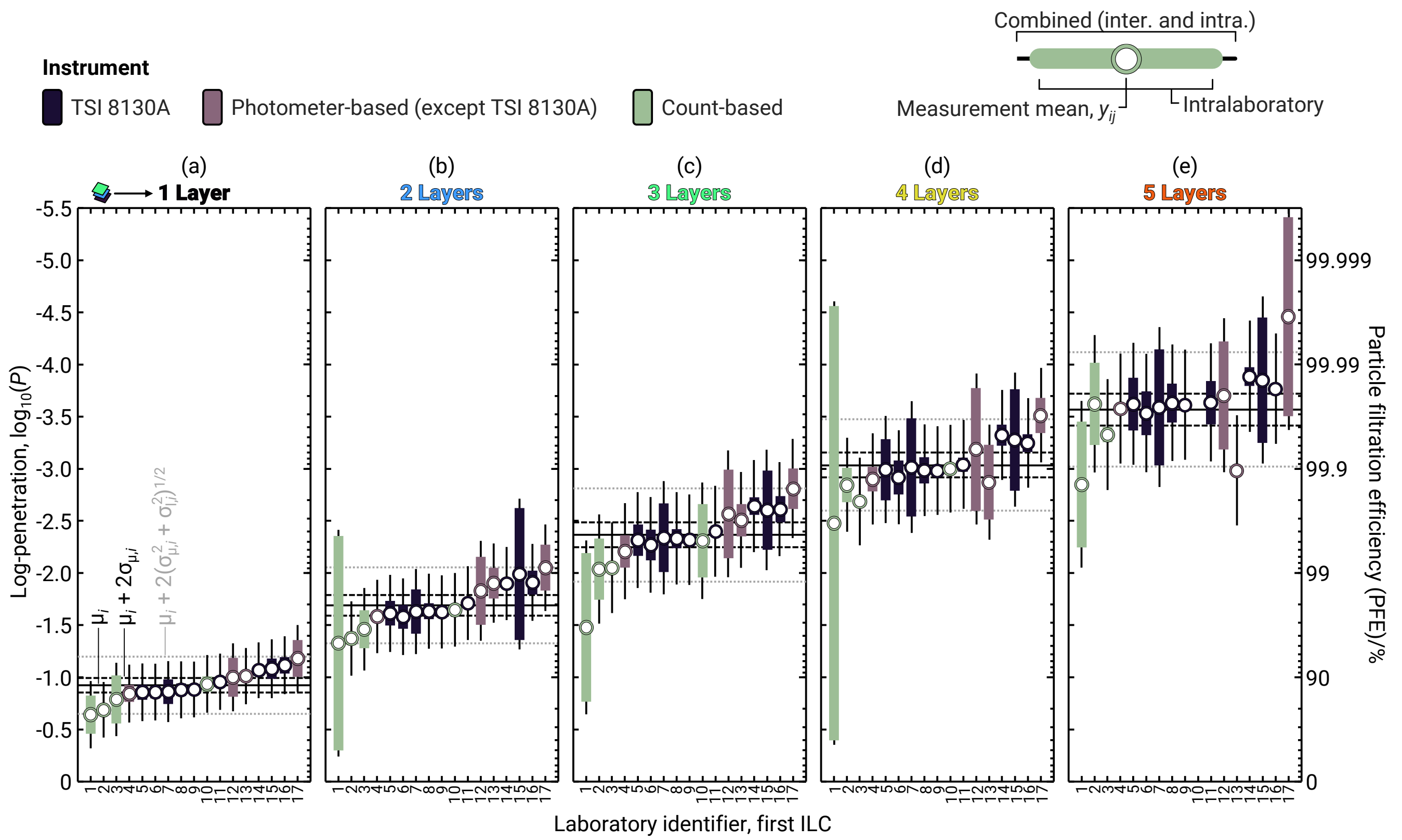

**Figure 1.** PFE results for an in increasing number of layers of Green Line Paper. Thick, coloured bars correspond to intralaboratory uncertainties for each laboratory, while thin, black whiskers correspond to combined uncertainty with added interlaboratory uncertainties. Horizontal rules, from the center out, correspond to the consensus value ($\mu_i$) (solid lines), intervals for uncertainties in the consensus value ($\sigma_{\mu,i}$) (dashed lines), and combined consensus and interlaboratory uncertainties ($[\sigma_{\mu,i}^2 + \sigma_{l,i}^2]^{1/2}$) (dotted lines). All uncertainties correspond to $k$ = 2.

Dark uncertainties (related to reproducibility) contributed significantly to the overall uncertainties across all of the subcomponents (see also Figure 8 later in this work). These contributions made up a larger proportion of the combined uncertainties at low PFE, in part as the repeatability was higher after the log-penetration transformation. Clear structure is visible in the results, with laboratories that measured a lower PFE for a single layer also typically measuring lower than the other laboratories when the material was layered. This is further indicative of structured biases between the laboratories for this type of material.

There is substantial scatter with respect to the different instrument types, though the mass-based PFE computed from count-based measurements tended to be lower. This observation could be related to a combination of remaining differences between scattering- and mass-based filtration efficiency, where scattering-based filtration tends to be higher [11]. Alternatively, this could stem from limitations in the existing verification procedure applied to the photometer instruments. The precise reason requires further study, in particular to quantify the latter contribution.

Figure 2 plots log-penetration against the pressure resistance for all of the laboratories, alongside isolines of constant quality factor, per Eq. (1). For each number of layers, there is a distinct central clustering of data (solid circles), with some other results that clearly deviate from this trend. For this reason, outlier detection is applied to the data prior to plotting, with open symbols corresponding to removed points. Deviant points typically correspond to outliers in the pressure resistance measurement, which could be removed using an aggressive outlier treatment (here, using a generalized extreme Studentized deviate test with a threshold factor of 1 instead of the default 0.6 used for log-penetration). This suggests that pressure resistance measurements could be improved, though this work does not address that specific issue. Note that the outliers in terms of log-penetration are identified with crosses in Figure 1. Within the central cluster of data, there is clear covariance between the measurements, with higher PFE for those results with a relatively high pressure drop. This is largely consistent with what we know about mechanical filtration, wherein an increase in the pressure drop is associated with thicker or denser materials, translating to a higher filtration. Overlaid ellipses for these central data clusters shown in Figure 2 correspond to $k = 2$ covariance following the outlier removal procedure.

Layering results in a trend of increasing filtration along a line of similar quality factor. This is consistent with expectations, where mechanical filtration increases according to an exponential with the thickness of the same material [5]. At the same time, the pressure resistance of the materials increases roughly linearly. There is some reduction in the quality factor with an increasing number of layers, decreasing from roughly 5.5 to 4.5. This is more evident when plotting with respect to the quality factor directly, Figure 2b. Plotting this way also shows that the spread in the quality factor reduces with increasing pressure drop and number of layers, consistent with observations for cotton fiber poplin weave in a lightweight flannel by Zangmeister et al. [19] albeit across different laboratories here. This is in part due to averaging of random variations in the underlying material over multiple layers. However, the standard error does not scale with the square root of the number of layers, rather saturating for more layers. This suggests a combination of the aforementioned effect with increasing counting errors (see also Figure 3 and surrounding discussion). Overall, despite the small decline in quality factor, this is a secondary effect, such that layering can be reasonably *approximated* using a constant quality factor, at least for this material.

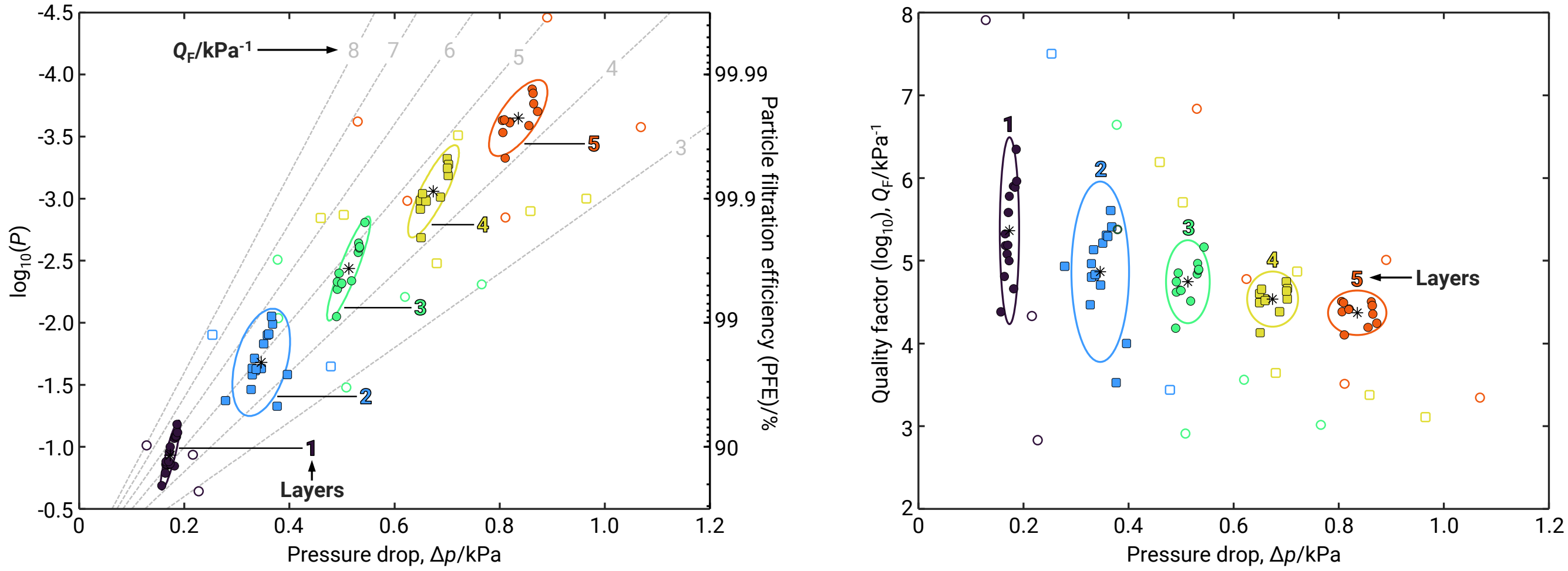

**Figure 2.** Relationship between the (a) logarithm of the penetration and (b) quality factor (from $\log_{10}$-penetration) and the pressure drop when layering Green Line Paper. Points correspond to the mean for each laboratory and test. Grey, dashed lines correspond to line of constant quality factor, $Q_F$, using logarithm base-10, which are linear in this space. Ellipses correspond to the covariance in non-outlier data ($k$ = 2). Solid symbols correspond to the central cluster, while open symbols correspond to the broader available data including outliers.

Figure 3 shows the size-resolved particle filtration for various numbers of layers of Green Line Paper measured using the PFEMS system [9, 10]. This system makes use of a pair of scanning mobility particle sizers (SMPSs), which classify particles according to their mobility diameter and then counts them to form distributions. Taking the ratio of particle count upstream and downstream of the sample allows for an assessment of the particle penetration as a function of particle size. Increasing noise in the size-resolved filtration as the PFE increases stems from a combination of low downstream counts and the log-penetration transformation, revealing visible single particle count or digital discretization noise in those signals. Noise at the edges of the curves representing the mobility diameter distributions correspond to cases where there was not a sufficient number of particles in the upstream distribution to allow for a robust measurement of particle filtration, where error propagation predicts a rapid expansion in measurement uncertainties [14].

The most penetrating particle size (MPPS) is relatively consistent across the different layers, again consistent with previous observations [19, 24] and simple models [5]. Further, the log-penetration at the MPPS appears to increase roughly linearly with the number of layers, such that the curves are evenly spread out vertically in the plot. This would correspond to a constant quality factor regardless of the number of layers, again consistent with simple models [5]. Plotting the quality factor, Figure 3b, shows that the size-resolved curves indeed collapse. In fact, a constant quality factor model seems to be more appropriate for the size-resolved PFE than for the PFE integrated over this size distribution shown in Figure 2. This is not unexpected. The size-resolved filtration is a more fundamental measurement in that it does not integrate over a range of filtration mechanism with different size effects. This observation further affirms that interfacial effects between the layers are secondary for these materials.

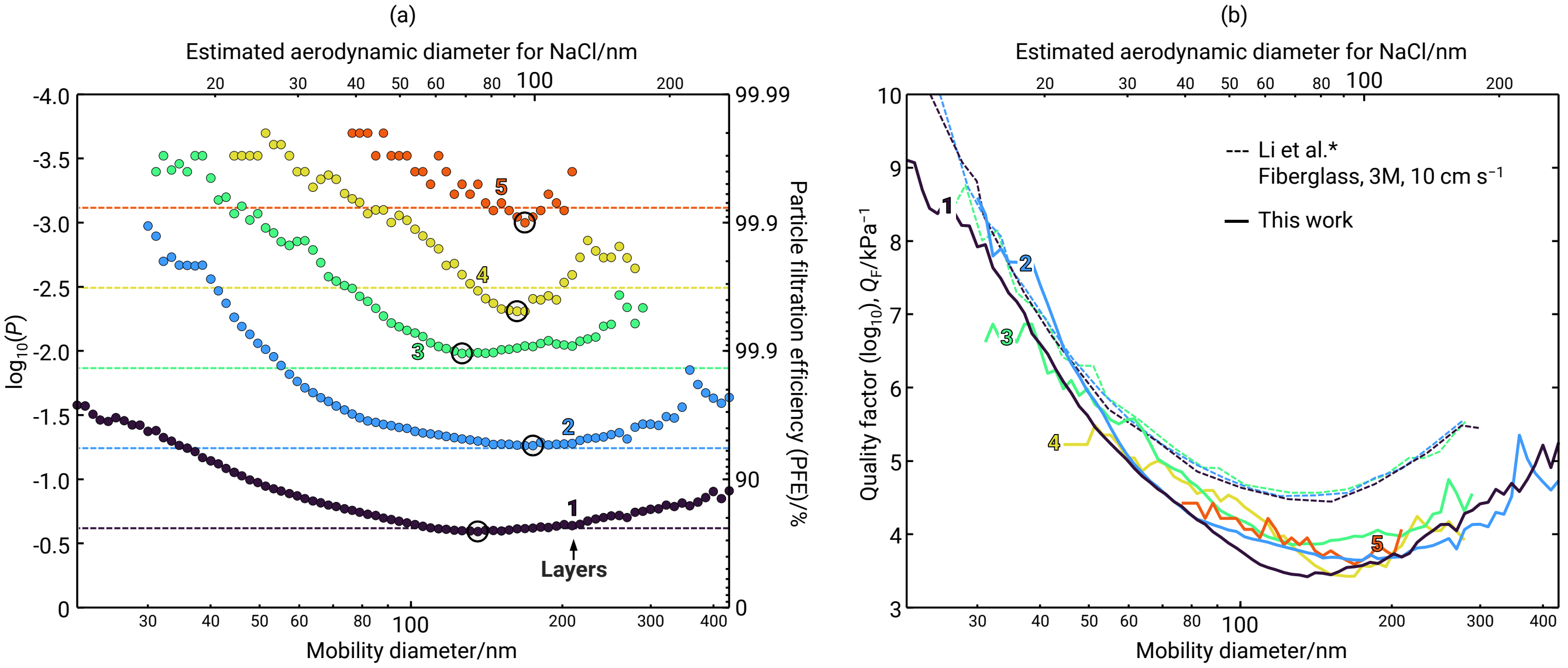

**Figure 3.** Size-resolved (a) particle filtration and (b) quality factor for various layers of Green Line Paper measured using upstream and downstream scanning mobility particle sizers. Upper axis gives an estimate of the corresponding aerodynamic diameter for the sodium chloride particles. In (a) the open circles correspond to the most penetrating particle size. Dashed lines in (b) correspond to digitized data from Li et al. [1], specifically the 3M data set at a face velocity of 10 cm s$^{-1}$. (*) Note that quality factors are computed by assuming (1) the pressure drop at 10 cm s$^{-1}$ is double that at 5 cm s$^{-1}$ and (2) that the pressure scales linearly from the value for a single layer. As such and unlike the data from this work, lines only assess the effect of penetration on the quality factor.

Quality factors measured here are also consistent with results from Li et al. [1], with the data for the fiberglass material as measured by 3M at a face velocity of 10 cm s$^{-1}$ also shown in Figure 3b. Given that pressure resistance was only reported at 5 cm s$^{-1}$ and only for a single layer, calculation of quality factors requires the assumption that the pressure resistance was a linear function of both the flow rate and number of layers. The resultant quality factors collapse onto a single curve, sharing a similar MPPS and overall trend in the size-resolved quality factor as the Green Line Paper measured in this work. Note that assuming the pressure drop is linear with the number of layers would reduce variability in the quality factor (given the functional form of the quality factor). Even so, the collapse of the data onto a single curve reinforces the sentiment that interfacial effects between the layers are insignificant for fiberglass. This results also appears to extend to the other fiberglass data sets reported by Li et al. [1]. Applying the same treatment to the data for flannel from Zangmeister et al. [19] adds further support to this conclusion, this at substantially lower quality factors (a minimum of $Q_F$ ~ 1.3 kPa$^{-1}$, noting the difference in the base of the log) and with some minor non-linearities in the pressure resistance that prevented full collapse (though these effects were small at less than 10 %). The electret materials considered by Li et al. [1], by contrast, did demonstrate decreasing quality factor with an increased number of layers, suggesting that interfacial effects may be significant for electret materials.

## 3.2 Candidate reference materials

PFE results for the candidate reference materials are shown in Figure 4, where dark, interlaboratory uncertainties were found to be significant across all of the materials tested and are dominant in all cases. Much of the expansion of dark uncertainties is associated with several measurements consistently above the central data cluster (these potential outliers are detected using GESD and are identified with crosses in Figure 4), which would add some skew in the underlying distribution of laboratory effects and is not fully captured in the chosen analysis method. These potential outliers expand the dark uncertainties by factors of 1.4 – 2.6. Test 6 (Figure 4f) is an exception, where there was sufficient spread in the data that no outliers were identified. The central data cluster after outlier removal was reasonably described by a normal distribution.

Again, substantial scatter was observed with respect to instrument type. However, in this case, the count-based instruments are also interspersed with the other results. This is roughly consistent with previous observations across a broader range of materials spanning to lower PFEs [11].

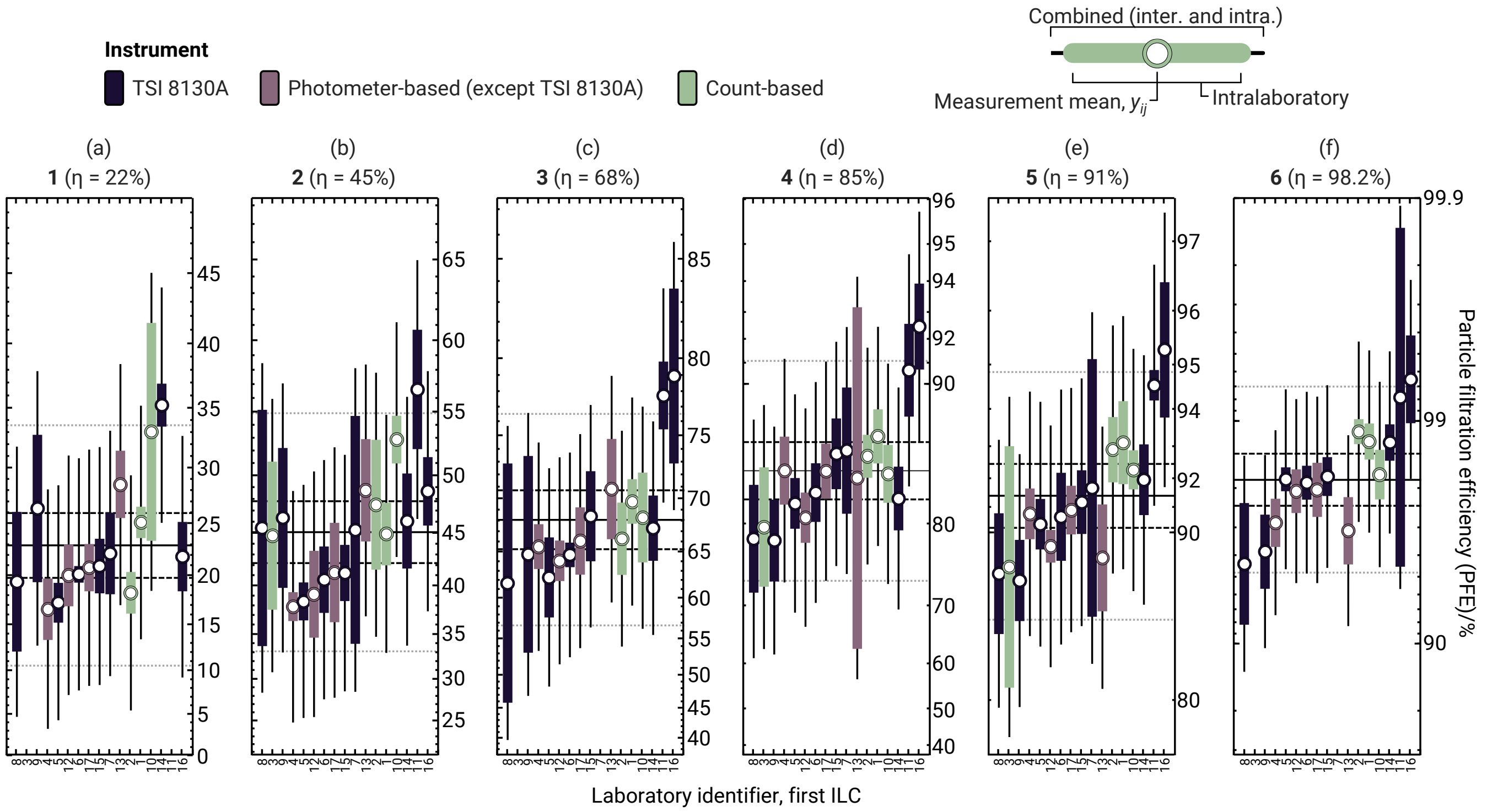

**Figure 4.** PFE results for the candidate reference materials with increasing consensus PFE, noting that the *y*-axes have different ranges across the panels. Laboratories are reordered in terms of the median across all of the panels. As with Figure 1, thick, coloured bars correspond to intralaboratory uncertainties for each laboratory, while thin, black whiskers include added interlaboratory uncertainties. Horizontal rules, from the center out, correspond to the consensus value (solid lines), intervals for uncertainties in the consensus value (dashed lines), and combined consensus and interlaboratory uncertainties (dotted lines). All uncertainties correspond to $k = 2$.

Figure 5 shows the relationship between pressure and log-penetration for these materials, following application of an outlier detection procedure, consistent with Figure 2. Results for each charge state generally follow lines of constant quality. Unlike Figure 2, the current results show this trend with increasing basis weight rather than layering of materials, showing that an analogous effect seems to apply between these two scenarios. This again suggests that interfacial effects are secondary to basis weight/layering across a broader set of materials (though not necessarily all).

Correlation between the pressure drop and penetration measured for a single material by the laboratories is less pronounced than in the Green Line Paper (manifesting as round circles instead of titled ellipses in Figure 5). This could be due to the lower pressure drop, resulting in larger coefficient in variation in the pressure drop and more scatter in that dimension. Alternatively, this may indicate that, unlike Green Line Paper, differences between samples of the material differ more in their makeup rather than simply having variability in the basis weight and/or thickness. It is worth noting that the uncharged case in Figure 5 also exhibits substantial scatter in pressure drop, indicating that the scattering in the pressure drop is independent of any charging effects, as expected. At the same time, it is also worth noting that scatter in the log-penetration dimension increases at higher charge states, which is likely an indication of the effect of variability in the applied charge expanding the uncertainties. The precise reason is difficult to discern and will be convolved with differences in the instruments and their calibration.

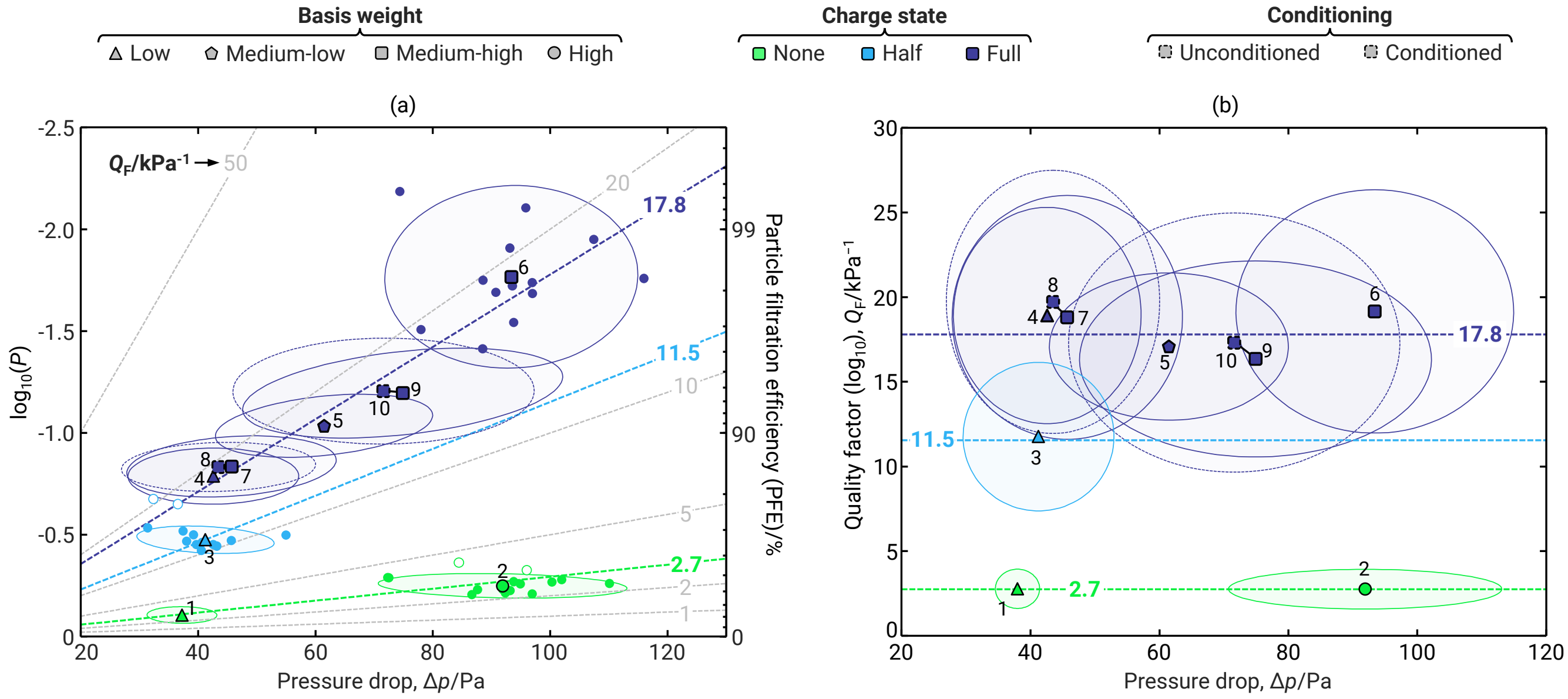

**Figure 5.** Relationship between the (a) logarithm of the penetration and (b) quality factor and the pressure drop across the candidate reference material. Results span different basis weights and charge states, though materials sharing a charge state had very similar quality across different basis weight categories and realizations. Select cases (2, 3, and 6) have original data overlaid on the plot, with open symbols corresponding to outlier data. Other cases have ellipses representing $k$ = 2 intervals on the covariance between the log-penetration and pressure drop.

## 3.3 Respirator tests

Both the final component of the first ILC and the entirety of the second ILC made use of respirators that were supplied to participants. Figure 6 shows the PFE results across the different measurements, which considered (a-b) the initial and minimum (while loading the material according to TEB-APR-STP-0059 [6]) and (c-e) the effect of conditioning and mounting direction (i.e., inhalation versus exhalation).

For the first ILC, repeatability within a laboratory was a good surrogate for the reproducibility across the laboratories in most cases. In other words, for the reporting laboratories, measurements seem to be reproducible both in terms of the initial and minimum PFE. Further, those laboratories with the least amount of interlaboratory variability also tended to have measurements close to the consensus value.

For the second ILC, an identical respirator was used across all three tests and an initial PFE was reported. Little structure was observed between measurements by the laboratories, with a random ordering in terms of PFE across the different panels (i.e., on average, no lab consistently measured below the others). Dark uncertainties were more significant for these respirators, driven in large part by laboratories that had repeatable measurements but are located at the periphery of the data (e.g., laboratory 1 in Figure 1c).

The use of an identical respirator necessitates that the only difference between the cases should stem from the change in conditioning and/or mounting direction. No statistically significant differences were observed in the consensus value between these different scenarios. Results were more varied for the exhalation case. This could indicate challenges associated with consistent mounting the respirators to face the other direction, which is a non-conventional setup. Yet, despite this reduction in reproducibility, the consensus PFE was not affected by mounting direction.

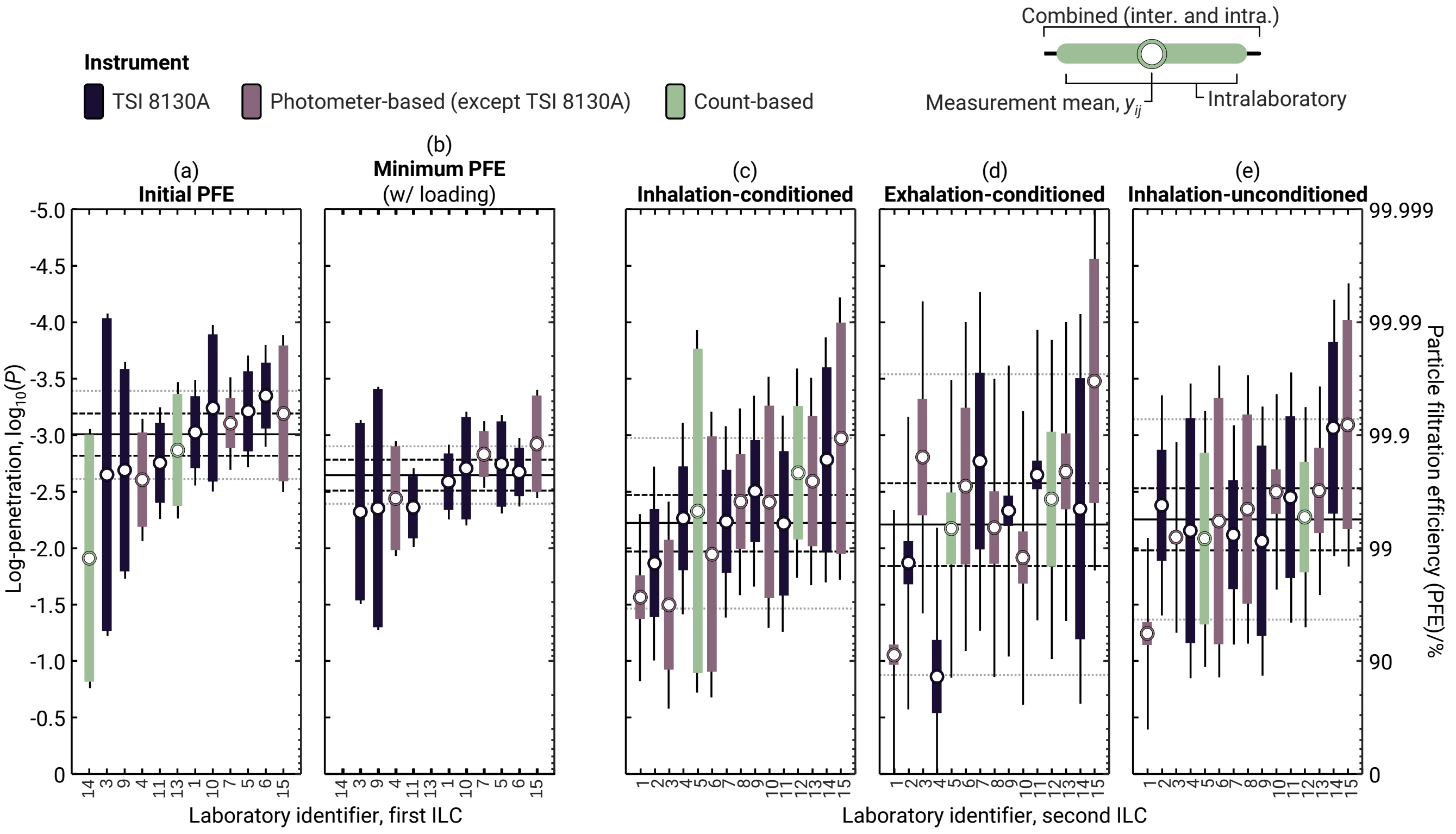

**Figure 6.** PFE results for testing respirators across (a-b) the final component of the first ILC and (c-e) the second ILC. Vertical axis is identical across all of the panels. For each ILC, the laboratories are sorted according to the first panel, that is (a) for the first ILC and (c) for the second ILC. The laboratory identifiers are not the same across the two ILCs (i.e., the laboratories in (a) and (b) are not the same as (c – e)). As with Figure 1, thick, coloured bars correspond to intralaboratory uncertainties for each laboratory, while thin, black whiskers correspond to added interlaboratory uncertainties. Horizontal rules, from the center out, correspond to the consensus value (solid lines), intervals for uncertainties in the consensus value (dashed lines), and combined consensus and interlaboratory uncertainties (dotted lines). All uncertainties correspond to k = 2.

## 3.4 Summary of uncertainty and its decomposition

Figure 7 summarizes the combined uncertainties across all of the samples and as a function of PFE. Uncertainties trended similarly across all of the samples, increasing in a relative sense (for log-penetration) with increasing PFE but decreasing significantly in an absolute sense. Thus, while relative uncertainties ranged from 11 – 30 % of the nominal value of log-penetration, corresponding 95 % credible intervals on the PFE vary from [10, 34] for the lowest PFE to [99.886, 99.994] for the highest PFE. The respirators from the second ILC were a bit of an exception to the overall trends, with larger uncertainties in particular for the exhalation case. This larger span was driven almost entirely by a few potential outliers, as evident in Figure 6d. Quality checks should be recommended for these laboratories.

Figure 8 summarizes the relative intralaboratory (repeatability), interlaboratory (dark, reproducibility), and consensus uncertainty contributions to the overall variance. Uncertainties in the consensus value were minimal for all of the measurements. In all but two respirator subcomponents, the majority of the combined uncertainties corresponded to dark, interlaboratory variability. Much of the observed dark uncertainties stem from what *might* be considered outlier laboratories. The removal of outliers by a GESD procedure gives an indication of the degree to which these stray measurements drive up dark uncertainties in the present analysis, with the leftover dark uncertainties shown as slightly darker, teal dashed border bars in Figure 8. The respirators are particularly prone to this effect. This is either an indication of the way overall uncertainties scale or that better controls are required on laboratory measurements to either (1) update the distribution used to describe the measurements or (2) improvements to quality control to ensure robust measurement across laboratories.

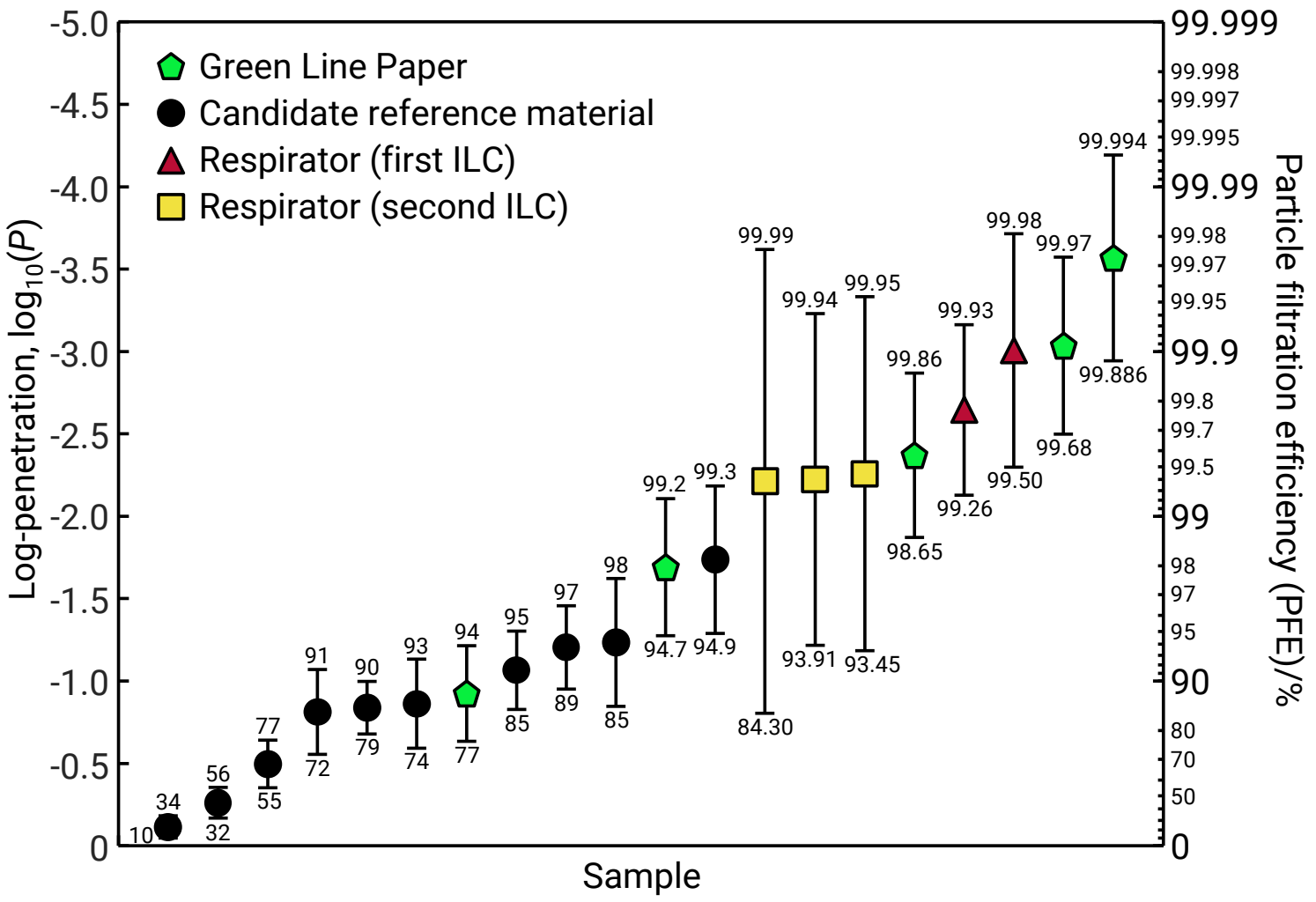

**Figure 7.** Consensus value with corresponding uncertainties ($k$ = 2) across all of the samples. Symbols and colors denote the different ILC subcomponents.

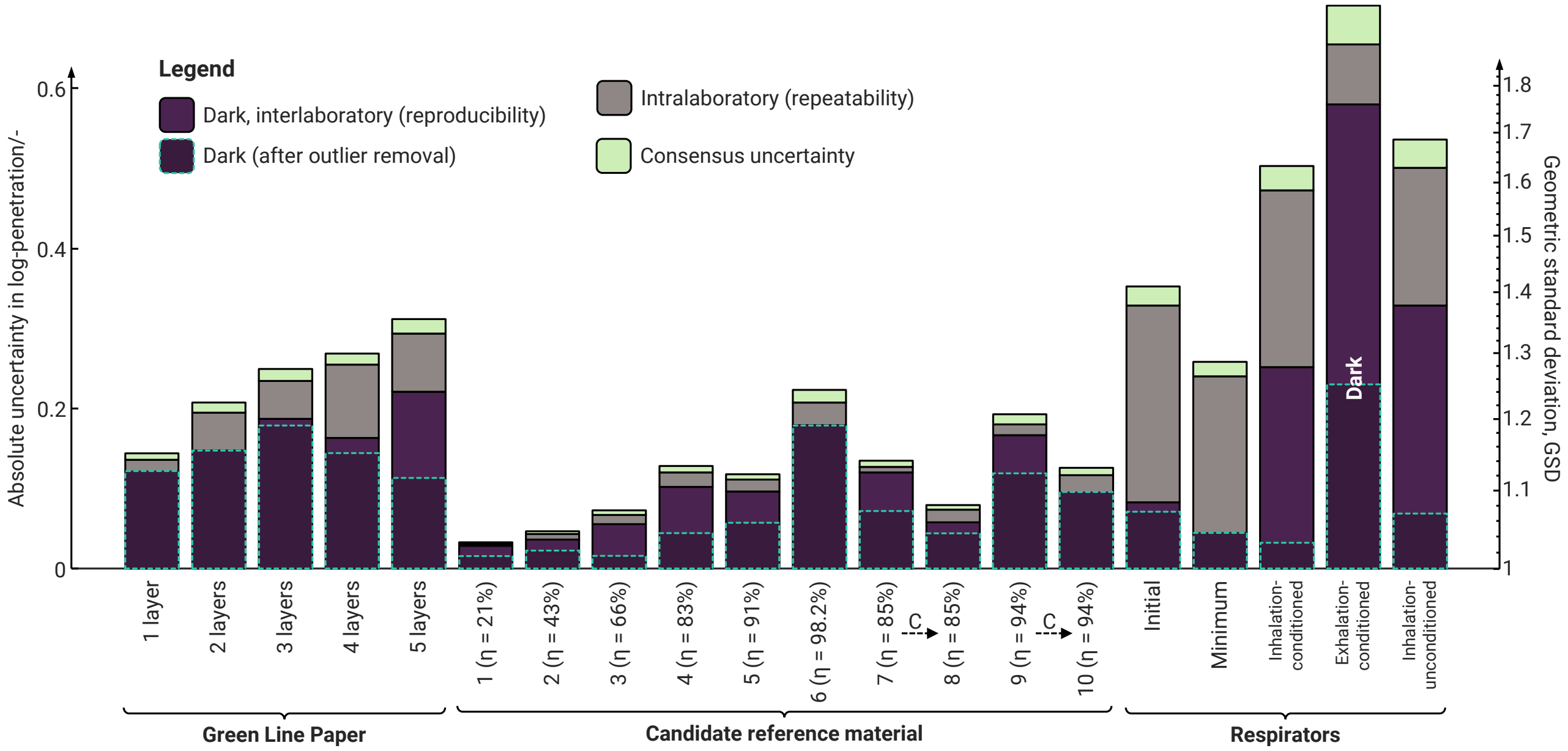

**Figure 8.** Absolute uncertainty and its decomposition for the PFE across all of the measurements. Lowest portions of the bars correspond to dark, interlaboratory contributions, followed by intralaboratory variability, and finally uncertainties in the consensus value at the top. Intralaboratory contributions are averaged over the various laboratories. Note that candidate reference materials 8 and 10 were conditioned version of materials 7 and 9, respectively.

## 4. Conclusions

This work has quantified the interlaboratory variability of particle filtration measurements for respiratory protection, spanning a range of materials for the conditions typical of testing respirators. Dark, interlaboratory variability was the dominant source of uncertainty in a vast majority of measurements. Overall uncertainties in the measurements varied from 10 % – 30 % of the nominal value of the log-penetration. Further investigation is required to understand and minimize the effect of both intralaboratory and interlaboratory (dark) differences in the measurements. Particular focus should be placed on the dark uncertainties, including studies of the procedures used to verify instrument function (e.g., used to constrain the challenge aerosol size distribution and number concentration as well as verifying the optical particle counters) and the persistence of outliers in these measurements, which contribute significantly to the dark uncertainties.

The uncharged Green Line Paper showed a linear relationship between the number of layers and log-penetration, consistent with simple multiplicative penetration through each layer, without any indication that interfacial effects are significant. Evidence for this approach is relatively clear in the size-resolved particle filtration, with some minor secondary effects when the particle filtration is integrated over a size distribution (for mass-based PFE). Similar results were observed when a controlled level of charge was applied to the candidate reference material. As such, this simple model and the traditional quality measure [5] should be a reasonable surrogate in terms of predicting the effect of layering and basis weight when the structure of the material does not otherwise change.

## Acknowledgements

This work was supported by the Pandemic Response Program at the National Research Council Canada and the Public Health Agency of Canada. We would also like to acknowledge participation by members of the Canadian PPE Laboratory Network and Australian laboratory network, including all the practitioners at the individual organizations that participated. The authors acknowledge the Canadian Association for Personal Protective Equipment Manufacturers (CAPPEM) for supplying some of the materials.